\documentclass[12pt]{article}

\usepackage{theorem,amssymb,amsbsy,latexsym,amsmath,bbm,epsfig,psfrag}

\theoremstyle{plain}

\newcommand{\delx}{\partial_x}

\newcommand{\half}{\mbox{$\frac12$}}
\newcommand{\quart}{\mbox{$\frac14$}}

\newcommand{\sign}{\,{\rm sign}} 

\newcommand{\nd}{{\phantom *}} 
\newcommand{\vx}{{\mbf x}}
\newcommand{\vn}{{\mbf n}}

\newcommand{\eps}{\varepsilon}
\newcommand{\mbf}[1]{{\boldsymbol {#1} }}
 
\newcommand{\cH}{{\cal H}}
\newcommand{\cQ}{{\cal Q}}
\newcommand{\cE}{{\cal E}} 
\newcommand{\ga}{\gamma} 
\newcommand{\ee}{{\rm e}}
\newcommand{\ii}{{\rm i}}

\newcommand{\Ref}[1]{(\ref{#1})}

\newcommand{\R}{{\mathbb R}}

\newcommand{\sct}[1]{\nopagebreak\noindent {\bf {#1}.}}

\begin{document}
\begin{flushright}
\today
\end{flushright}
\vspace{.4cm}

\begin{center}

{\Large \bf Singular factorizations, self-adjoint extensions, and applications
to quantum many-body physics}

\vspace{1 cm}

{\large Edwin Langmann$^*$, Ari Laptev$^{**}$ and Cornelius
Pauf\/ler$^{***}$}\\
\vspace{0.3 cm} {\em $^*$ Mathematical Physics, KTH Physics, AlbaNova,
SE-106 91 Stockholm, Sweden}\\ {\em $^{**}$ Department of Mathematics,
KTH, SE-100 44 Stockholm, Sweden}\\ {\em $^{***}$ Theoretische
Elementarteilchen-Physik, Institut f\"ur Physik,
Johannes-Gutenberg-Universit\"at, D-55099 Mainz, Germany}\\

\end{center}

\begin{abstract}
We study self-adjoint operators defined by factorizing second order
differential operators in first order ones.  We discuss examples where
such factorizations introduce singular interactions into simple
quantum mechanical models like the harmonic oscillator or the free
particle on the circle. The generalization of these examples to the
many-body case yields quantum models of {\em distinguishable} and
interacting particles in one dimensions which can be solved explicitly
and by simple means.  Our considerations lead us to a simple method to
construct exactly solvable quantum many-body systems of
Calogero-Sutherland type.

\end{abstract}

\sct{1. Introduction}

\noindent In this paper we define and study peculiar self-adjoint
extensions of simple differential operators like the one defining the
quantum mechanical harmonic oscillator. We also find the explicit
solution of some of the resulting quantum mechanical models and
discuss their physical interpretations.  These examples provide
amusing illustrations that the mathematical theory of Schr\"odinger
operators \cite{Kato,RSI,RSII,BS}, and in particular the distinction
between formally hermitian and self-adjoint operators, can have
important physical consequences even in elementary quantum mechanics.
Moreover, these examples allow a straightforward generalization to the
case of an arbitrary number of particles, and the resulting many-body
models describe strongly interacting systems which can be solved by
elementary means. These models include, as special cases, the ones
studied by Girardeau \cite{G}, which have received considerable
interest recently since it has become possible to realize and study
them experimentally \cite{Nature}.  We stress that these models can be
solved even in the general case of {\em distinguishable} particles,
and that they can be studied by much simpler methods than other
well-known integrable quantum many-body systems
\cite{LL,Y,C,S}. Moreover, our considerations lead us to a remarkably
simple and general approach to exactly solvable quantum many-body
systems of Calogero-Sutherland type \cite{C,S,OP} which seems new.

The starting point for this work was a paradox for the differential
operator $-\delx^2 + x^2$ defining quantum harmonic oscillator,
$-\infty <x <\infty$ and $\delx := \partial/\partial x$. The paradox
arises if one confronts the formal factorization
\begin{equation} 
\label{2} 
-\delx^2 + x^2 =\left(\delx -x+\frac1x\right)
\left(-\delx -x+\frac1x\right) +3 \equiv Q^*Q+3  
\end{equation}
and $Q^*Q\geq 0$, with the well-known fact that the groundstate
eigenvalue of the harmonic oscillator Hamiltonian $H=-\delx^2 + x^2$
is $1$. It is interesting to note that a variant of this paradox was
already discussed by Calogero in 1969; see \cite{C0}, second paragraph
after Eq.\ (3.24).

The paradox above is only one of many, and they are resolved by
observing that each such factorization gives rise to a different
quantum mechanical Hamiltonian defined by the Friedrichs
extension. These examples have in common that a simple differential
operator with an obvious factorization also allows for alternative
factorizations where singularities are introduced, and this implies a
restriction on the domain of functions on which the resulting
self-adjoint operator is defined. This restriction has a natural
physical interpretation as strong repulsive interaction.

We will present several examples leading to interesting quantum
mechanical model with additional interactions and which are exactly
solvable in the sense that all eigenfunctions and eigenvalues can be
found explicitly and by simple means.\footnote{To illustrate what we
mean by this we recall that the spectrum of the Hamiltonian defined by
the differential operator $-\delx^2$ on all $C^\infty$-functions on an
interval and general boundary conditions is given by a transcendental
equation in general which can be solved only numerically. These
examples are not exactly solvable in this sense.}  In particular, for
each non-negative integer $n$ we will find a quantum mechanical
Hamiltonian $H_n$ which is defined by the differential operator of the
harmonic oscillator differential operator $-\delx^2+x^2$ but is
bounded from below by $2n+1$.  Moreover, in addition to the standard
harmonic oscillator $H=H_0$ also the Hamiltonian $H_1$ defined by the
factorization in Eq.\ \Ref{2} is exactly solvable, while the other
$H_{n\geq 2}$ are only partially solvable, i.e., only a finite number
of eigenstates and eigenvalues can be found exactly. Two other
examples are provided by the differential operator $-\delx^2 -
\ga(\ga+1)\cosh^{-2}(x)$ often referred to as P\"oschl--Teller
potential, and the differential operator $-\delx^2$ for $-\pi < x
<\pi$ describing a free particle on the circle. In the latter example
we will show that, for every $\alpha>0$, the factorization
$$
-\delx^2 = \bigl( \delx + \alpha \cot(\alpha x)\bigr)\bigl(
-\delx + \alpha \cot(\alpha x)\bigr) + \alpha^2
$$
defines a unique self-adjoint operator $H_\alpha$ bounded from below
by $\alpha^2$. Moreover, this Hamiltonian is exactly solvable if and
only if $\alpha$ is integer or half integer.

While we believe that these examples are of pedagogical interest, our
first main result is that much of what we say can be naturally
extended to the many-body case.  This yields many-body systems
describing distinguishable particles moving in one dimension and
interacting through a strong local two-body interactions, and also
these models can be solved exactly and by simple means.

Our considerations described above lead us, somewhat unexpectedly, to
our second main result which is on Calogero-Sutherland type systems:
As we will see, all our one-particle examples mentioned above are
related to Schr\"odinger operators $H=-\delx^2 + V(x)$ which have
exact eigenfunctions of the form
\begin{equation}
\psi_n(x) = P_n(\eta(x))\, \ee^{-w(x)} ,\quad n=0,1,2,\ldots ,
N_{max},\label{psin}
\end{equation} 
where $P_n$ is a polynomial of order $n$, $\eta$ and $w$ are
real-valued $C^2$ functions, and $N_{max}\leq \infty$. These quantum
mechanical models are exactly solvable in the sense that these
eigenfunctions $\psi_n$ and the corresponding eigenvalues $E_n$ can be
found explicitly.  While we will restrict our discussions to our three
examples: harmonic oscillator, P\"oschl-Teller potential, and free
particle on the circle, it is important to note that there are many
more: in Table~\ref{table1} we list the well-known examples associated
with the classical orthogonal polynomials, but even this list is not
complete. Our second main result is a surprisingly simple argument
showing that, to {\em any} such model, the corresponding quantum-many
body given by the Schr\"odinger operator
\begin{equation}
H_N = \sum_{j=1}^N \Bigl( -\partial_{x_j}^2 + V(x_j) \Bigr) +
\lambda(\lambda-1) \sum_{1\leq j<k\leq N} W(x_j,x_k)\label{HN}
\end{equation}
and the following particular two-body interaction
\begin{equation}
W(x,y) = \frac{\eta'(x)^2+\eta'(y)^2}{\left(\eta(x)-\eta(y)\right)^2}
 + \frac{ 2\eta'(x)w'(x) -\eta''(x)
 -2\eta'(y)w'(y)+\eta''(y)}{\eta(x)-\eta(y)}-W_0 \label{W}
\end{equation}
defines an exactly solvable quantum many-body system;
$\eta'(x):=\delx\eta(x)$, and $W_0$ is an arbitrary constant which can
be fixed to some convenient value. More specifically, this arguments
leads to a simple proof of the following:

\bigskip

\noindent {\bf Proposition:} {\it Let $H=-\delx^2 + V(x)$ be a
self-adjoint Schr\"odinger operator on $L^2([a,b])$, $-\infty\leq
a<b\leq \infty$, with exact eigenfunctions as specified in Eq.\
\Ref{psin} {\em ff}.  Then the corresponding many-body Schr\"odinger
operator $H_N$ in Eqs.\ \Ref{HN}--\Ref{W} defines a quantum-many body
systems with the following exact ground state,
\begin{equation}
\Psi_0(x_1,\ldots,x_N) = \ee^{-\sum_{j=1}^N w(x_j) } \prod_{1\leq j<k\leq N}
\Bigl( \eta(x_j) - \eta(x_k) \Bigr)^\lambda ,\label{Psi0} 
\end{equation}
provided $N<N_{max}+1$ and $\lambda>0$ is such that $\Psi_0$ is
square-integrable. The corresponding eigenvalue is determined by the
one-particle eigenvalues $E_n$ as follows,
\begin{equation} \cE_0 = \lambda^2\sum_{n=0}^{N-1}(E_n-E_0) +
NE_0-\frac12\lambda(\lambda-1)N(N-1)W_0. \label{cE0}
\end{equation}
}
\bigskip

\noindent (A concise proof can be found in the last paragraph of Section~6.)

\bigskip

\begin{table}[htbp]
  \begin{center}
    \begin{tabular}{c|cccc}
      & Name & $\psi_n(x)$ & $V(x)$ & $E_n$ \\ 
      \hline I & Jacobi &
      $(\sin\frac{x}2)^\alpha(\cos\frac{x}2)^\beta
      P_n^{(\alpha,\beta)}(\cos x)$ &
      $\frac{4\alpha^2-1}{16\sin^2\frac{x}2} +
      \frac{4\beta^2-1}{16\cos^2\frac{x}2}$&
      $\left(n+\frac{\alpha+\beta+1}2\right)^2$\\
      II & Gegenbauer & $(\sin x)^\alpha C_n^{(\alpha)}(\cos x)$ &
      $\frac{\alpha(\alpha-1)}{\sin^2 x}$ & $(n+\alpha)^2$ \\  
      III & Chebyshev & $T_n(\cos x)$ $\phantom{\frac{\frac
      AB}{\frac AB}}$ & $0$ & $n^2$ \\
      IV & Laguerre & $\ee^{-x^2/2} x^{(\alpha+1)/2}
      L_n^{(\alpha)}(x^2)$ & $4n+2\alpha+2$ & $x^2 +
      \frac{4\alpha^2-1}{4x^2} $ 
    \end{tabular}
    \caption{Solutions of the Schr\"odinger equation
    $\left(-\partial_x^2+ V(x)-E_n\right)\psi_n(x)=0$ associated with
    classical orthogonal polynomials; adapted from Table~22.6 in
    \cite{AS}.}
    \label{table1}
  \end{center}
\end{table}

In fact, we found that it is possible compute all other bound states
and corresponding eigenvalues for all these systems exactly and in a
unified manner \cite{HL3}. As we will discuss in Section~7, special
cases of this result yield various well-known exactly solvable systems
of Calogero-Sutherland type \cite{C,S,OP,BF}, but there are other
interesting examples which we have not seen discussed in the physics
literature.

The plan of this paper is as follows. In the next section we resolve
in detail the paradox described above using the theory of Friedrichs
extensions. This provides a general recipe for finding other such
examples. In Section~3 we present our one-body examples, and in
Section 4 we discuss their physical interpretation.  In Section~5 we
discuss the extension of our results to the many-body case.  Our
method to find quantum-many body systems with exact groundstates is
explained in Section~6. We end with a few remarks in Section~7.

\sct{2. Factorizations and selfadjoint extensions}

\noindent We first recall some standard facts from the operator theory
in Hilbert spaces.  Let $A_0$ be a symmetric operator in a separable
Hilbert space ${\cal H}$ with scalar product $(\cdot, \cdot)$ (linear
in the second argument) and norm $\|\cdot\|$, and let ${\cal D}(A_0)$
be the domain of the operator $A_0$ which is a dense subset in ${\cal
H}$. We assume that $A_0$ is bounded from below, i.e., there exists a
constant $E_0\in {\mathbb R}$ such that
$$
(A_0 u,u)\ge E_0\|u\|^2, \qquad u\in{\cal D}(A_0).
$$
The sesquilinear form $a_0[u,v]= (A_0 u,v)$ defined for 
$u,v\in d[a_0] = {\cal D}(A_0)$
can be completed with respect to the norm defined by 
\begin{equation}
\label{sesqui}
a_0[u,u] +(E_0+1)\|u\|^2.
\end{equation}
Such a completion is denoted by $a$ and its domain by $d[a]$.  By
standard arguments (see for example \cite{BS}) there is a unique
selfadjoint operator $A$ corresponding to the closed form $a$ whose
domain ${\cal D}(A)\subset d[a]$. This operator is called the
Friedrichs extension of the operator $A_0$.

\medskip
\noindent
Let $V$ be a real function defined on open subset 
$\Omega\subseteq {\mathbb R}^N$, 
and let $H$ be a self-adjoint Schr\"odinger 
operator in $L^2(\Omega)$ with some boundary conditions.
\begin{equation}
\label{Schr} 
H u(x) =  -\Delta u(x) + V(x)u(x).
\end{equation}
This operators generates a quadratic form $h[u,u]$ defined on
functions from $d[h]$. We assume that $H$ has discrete spectrum
$\{E_n\}_{n=0}^K$, with the corresponding orthonormal in $L^2(\Omega)$
system of eigenfunctions $\{u_n\}_{n=0}^K$ where $K\leq \infty$ (i.e.\
$H$ may also have continuous spectrum).  It can be easily checked
directly that the differential expression $-\Delta + V(x)$ can
formally be factorized as
\begin{equation}
\label{factor}
 -\Delta + V = Q_n^*Q_n + E_n,
\end{equation}
where 
\begin{equation}
\label{Q}
Q_n=-\nabla_x + \frac{\nabla u_n}{u_n} .
\end{equation}
It is well-known that the ground state $u_0$ can be chosen strictly
positive and the factorization \eqref{factor} does not give any
problem. However, considering higher eigenfunctions we obtain
singularities in the right hand side of \eqref{Q} due to the zero set
$S$ of $u_n$, $S = \{x: u_n(x) = 0\}$.  In order to define the
selfadjoint operator corresponding to the operator $Q^*Q$ we begin
with the quadratic form 
$$
h_n[u,u] = \int_\Omega \Bigl| -\nabla_x u + \frac{\nabla u_n}{u_n} u
\Bigr|^2 \, {\rm d}^Nx,
$$
whose domain $d[h_n]$ consists of functions
$u\in d[h]$  such that $u(x)=0$ for $x\in S$.

The corresponding Friedrichs extension gives us a selfadjoint operator 
which we denote by $H_n$. Although  
\begin{equation}
\label{contr1} 
H_n  = Q_n^*Q_n + E_n = -\Delta + V
\end{equation}
coincides with $H= -\Delta + V$ defined by \eqref{Schr} regarding its
action on functions as a differential expression, its domain is very
different and consists of functions not only just satisfying the
boundary conditions of the original operator $H$ but being equal to
zero on $S$.  This automatically makes the domain $d[h_n]$ smaller
compared with the domain of the original operator $H$ and ``lifts up''
the spectrum.

In particular, this answers the question concerning the formal
inequality which follows from \eqref{contr1}, that is
$$
H  =  Q_n^*Q_n + E_n \ge E_n.   
$$ The equality in the right hand side of this latter equation is
wrong because the domains of the operator $H$ and $Q_n^*Q_n$ are
different and therefore $H\not= Q_n^*Q_n + E_n$ as Hilbert space
operators.

\sct{3. Some examples}

\noindent Here we shall give some simple examples where the above
construction can be worked out more explicitly.
 
\medskip
\noindent
{\bf 3.a.} Harmonic oscillator.

\smallskip
\noindent
Let $H$ be the operator of the harmonic oscillator on $\R$,
\begin{equation}
\label{harm}
H =-\delx^2 + x^2. 
\end{equation}
The eigenvalues and eigenfunctions of this operator are well-known and
equal to $E_n= 2n+1$, $n=0,1,\dots$, and
\begin{equation}
\label{psin1} 
\psi_n(x) = \cH_n(x) e^{-x^2/2},
\end{equation}
where $\cH_n$ are Hermite polynomials, 
$$ \cH_n(x) = (2^n n!)^{-1/2} \pi^{-1/4} (-1)^n e^{x^2}\delx^n
e^{-x^2}.
$$
As in \eqref{contr1} the operators $H_n$ can be defined via 
$$
H_n  = Q_n^*Q_n + E_n = \Bigl(\delx +
\frac{\psi_n^\prime}{\psi_n}\Bigr) \Bigl(-\delx +
\frac{\psi_n^\prime}{\psi_n}\Bigr) + E_n. 
$$
According to Section 2 the latter expression coincides with
$-\delx^2 + x^2$ as a differential expression but not as a
selfadjoint operator.  In particular, when $n=0$,
$$
\frac{\psi_0^\prime}{\psi_0} = -x, 
$$ 
and we obtain the standard factorization of the harmonic oscillator
$$
H_0= -\delx^2  + x^2 = \Bigl(\delx  - x\Bigr)
\Bigl(-\delx  - x\Bigr) + 1. 
$$
If $n=1$ then 
$$
\frac{\psi_1^\prime}{\psi_1} = -x + \frac{1}{x} , 
$$ 
and we obtain the factorization 
$$
H_1 = -\delx^2 + x^2 = \Bigl(\delx - x + \frac{1}{x}\Bigr)
\Bigl(-\delx - x + \frac{1}{x}\Bigr) + 3.
$$

It is interesting to note that not only $H_0$ but also $H_1$ defines an
exactly solvable system in the sense that all its eigenfunctions can be
computed explicitly: Using that the eigenfunctions $\psi_n(x)$ of $H_0$ in Eq.\
\Ref{psin1} {\em ff.} are even and odd for even and odd integers $n$,
respectively, we conclude that the eigenfunctions $u_n(x)$ of $H_1$ are given
by
\begin{equation} 
u_{2n+1}(x) = \psi_{2n+1}(x) ,\quad u_{2n}(x) = \psi_{2n+1}(x)
\sign(x),\quad n=0,1,2\ldots 
\end{equation}
with the usual sign function $\sign(x)$: the odd eigenfunctions of
$H_1$ are the same as for the standard harmonic oscillator, but the
even ones are very different. The spectrum of $H_1$ thus has
multiplicity two, $E_{2n}=E_{2n+1} = 4n+3$.

The other exotic harmonic oscillator Hamiltonians $H_{n\geq 2}$ are
only partially solvable: as discussed in Section~4 below, we can
determine the $(n+1)$-fold degenerate ground state of $H_{n\geq 2}$
explicitly, but it is not possible to compute its excited states
analytically.

\medskip
\noindent
{\bf 3.b.} P\"oschl--Teller potential.

\smallskip
\noindent
Let 
\begin{equation}
\label{HPT} 
H= -\delx^2 - \frac{\ga(\ga+1)}{\cosh^2x} 
\end{equation}
for $\ga>0$.  It is well-known that this operator has $[\ga]$
(= largest integer smaller or equal to $\ga$) square integrable
eigenfunctions $\psi_n$, $n=0,1,\ldots, [\ga]-1$, with corresponding
eigenvalues
\begin{equation} 
E_{n} = -(\ga -n)^2 .  
\end{equation}
These eigenfunctions are given by 
\begin{equation} 
\psi_{n}(x) = \cosh^{-\ga}(x)\sum_{0\leq s\leq n/2}
(-1)^s\frac{n!(\ga-n)!}{4^s(n-2s)!(\ga+s-n)!s!} \sinh^{n-2s}(x)
\label{PTn} 
\end{equation}
with $r! := \Gamma(r+1)$ for non-integers $r$, as can be checked by
straightforward computations.

In particular, the lowest eigenvalue of this operator is
$E_{0}=-\ga^2$ and the corresponding eigenfunction is
$\psi_{0}(x) = \cosh^{-\ga}(x)$.  Then
$$
\frac{\psi_0^\prime}{\psi_0} = -\ga\tanh(x)
$$
and therefore one can obtain the following factorization
$$
H_0= -\delx^2  - \frac{\ga(\ga+1)}{\cosh^2x} 
= \Bigl(\delx  
 -\ga\tanh(x)\Bigr)
\Bigl(-\delx  -\ga\tanh(x)\Bigr) - \ga^2
$$ defining the standard selfadjoint operator $H_0=H$ given by the
differential operator $-\delx^2+\ga(\ga+1)\cosh^{-2}(x)$. The second
eigenvalue is $E_{1} = -(\ga -1)^2$ and the corresponding
eigenfunction is given by $ \psi_{1}(x) = \cosh^{-\ga}(x)\sinh(x)$,
for $\ga>1$.  Then
$$
\frac{\psi_1^\prime}{\psi_1} = \coth(x) -\ga\tanh(x)
$$ 
which implies that 
\begin{eqnarray*}
H_1= -\delx^2 - \frac{\ga(\ga+1)}{\cosh^2x} &=& \Bigl(\delx + \coth(x)
-\ga\tanh(x)\Bigr)\\ && \times \Bigl(-\delx + \coth(x)
-\ga\tanh(x)\Bigr) -(\ga -1)^2
\end{eqnarray*}
which defines another selfadjoint operator $H_1$, and similarly for
$H_{n\geq 2}$. Again it is possible to compute all eigenfunctions and
eigenvalues of $H_1$ in terms of the ones of $H_0$, similarly as in
the harmonic oscillator case.

\medskip
\noindent
{\bf 3.c.} Free particle on the circle.

\smallskip
\noindent
This example is slightly more subtle. For arbitrary $\alpha>0$ we define a
self-adjoint operator on $L^2([-\pi,\pi])$ by the factorization
$$
H_\alpha = -\delx^2 = (\delx + \alpha \cot(\alpha x))
(-\delx + \alpha \cot(\alpha x)) + \alpha^2 
$$ which is obtained with the odd eigenfunctions $\sin(\alpha x)$ of
$-\delx^2$. The subtle point now concerns the correct boundary conditions
at $x=\pm \pi$. According to our general theory in Section~2, we need to start
with the self-adjoint operator $H=-\delx^2$ with boundary conditions such
that $\sin(\alpha x)$ is indeed its eigenfunction, and it is not difficult to
see that these conditions are as follows,
\begin{equation}
\label{SL-bc}
  u'(-\pi)=u'(\pi)\; \textrm{ and } \; u(\pi)-u(-\pi)= - 2\tan(\alpha\pi)
  u'(\pi) 
\end{equation}
for functions $u$ in the domain of $H$. Thus only for integers $\alpha$ this
corresponds to periodic boundary conditions, and only in this case can we
interpret $H$ as Hamiltonian of a particle on a circle. We conclude that
the eigenfunctions of $H_\alpha$ are defined by the same boundary conditions
at $x=\pm \pi$ together with the requirements that they vanish at all points
$x$ where $\sin(\alpha x)=0$.

\noindent If we interpret $H_\alpha$ as Hamiltonian of a particle moving on a
circle, then we not only have infinitely strong repulsive interactions at the
points $x$ where $\sin(\alpha x)$ vanishes but, in addition, a singular
interaction at the point $x=\pm \pi$ whose strength depends on $\alpha$ so
that it vanishes at integer values of $\alpha$ and becomes infinitely strong
if $\alpha$ is half integer.

\noindent
In particular, if $\alpha= 1/2$, then the operator
$$
H_{1/2} = -\delx^2 = \left(\delx + \half \cot(x/2) \right) \left(-\delx
+ \half \cot(x/2) \right) + \quart
$$
is defined on functions $u$ satisfying a Dirichlet boundary condition
at zero (from the singularity of $\textrm{cot}$ at zero) 
and Neumann boundary conditions at $-\pi$ and $\pi$, 
$$
u(0)=u'(-\pi)=u'(\pi) = 0 . 
$$

\noindent 

Our discussion above implies that, for integers and half-integers
$\alpha$, one can determine all eigenfunctions and eigenvalues of
$H_\alpha$ explicitly and by elementary computations, but otherwise
part of the spectrum is determined by a transcendental equation: the
eigenfunctions are determined by the conditions $u(k\pi/\alpha)=0$ for
integer $k$ and the boundary conditions in Eq.\ (\ref{SL-bc}), and the
latter are non-trivial unless $2\alpha$ is an integer.

\noindent For example, for $\alpha=1$ these eigenfunctions are
given by
$$
u_{\pm,n}(x) = \Theta(\pm x) \sin(nx)
$$ with $n=1,2,\ldots$. The corresponding eigenvalues are $n^2$ and
two fold-degenerate.  For $\alpha=1/2$ the explicit eigenfunctions are
$$
u_{\pm,n} = \Theta(\pm x) \sin((n+1/2)x/2) \quad n=0,1,2,\ldots  
$$ with corresponding two-fold degenerate eigenvalues $(n+1/2)^2$.

We finally mention that it is amusing to visualize the $\alpha$-dependence of
the singular interactions and the spectrum of the Hamiltonian $H_\alpha$.

\sct{4. Physical Interpretation}

\noindent We first give a physics argument providing an intuitive
resolution of the paradox mentioned after Eq.\ (\ref{2}).

The paradox arises due to the singularity in the factorization in Eq.\
(\ref{2}) at $x=0$. One can regularize this singularity and consider
the following family of Hamiltonians,
$$H^{(\eps)} =\left(\delx -x+\frac{x}{x^2+\eps^2}\right) \left(-\delx
-x+\frac{x}{x^2+\eps^2}\right) +3 , $$
with a regularization parameter $\eps>0$. By a straightforward
computation one finds
$$
H^{(\eps)} = -\partial_x^2 + x^2 + \frac{\eps^2(1+2\eps^2 +
  2x^2)}{(x^2+\eps^2)^2} . 
$$ Even though the second term vanishes for $\eps=0$, it is obviously
relevant in the limit $\eps\downarrow 0$: Recalling that
$\delta_\eps(x)= \eps/(x^2+\eps^2)\pi$ is a sequence of $C^\infty$
functions converging to the delta distribution as $\eps\downarrow 0$,
we see that this term gives an additional singular interaction of
infinite strength at $x=0$: formally, the second term converges to
$\lim_{\eps\downarrow 0} \left(\pi\delta_\eps (x)\right)^2 =
\pi^2\delta(0)\delta(x)=\lim_{g\to+\infty} g \, \delta(x)$. This
interaction forces the eigenfunctions of the Hamiltonian
$\lim_{\eps\downarrow 0} H^{(\eps)}$ to vanish at $x=0$. It is
plausible that this latter Hamiltonian is identical with the one
defined by the factorization in \Ref{2}, and this is indeed the case,
as discussed below.

\noindent As shown in the previous section, the eigenfunctions of
Hamiltonians $H=-\delx^2 + V(x)$ on $L^2([a,b])$, $-\infty\leq a<b\leq
\infty$, defined by the factorization
$$ H_n = -\delx^2 + V(x) = \left( \delx + u_n'(x)/u_n(x)\right) \left(
-\delx + u_n'(x)/u_n(x)\right) +E_n
$$
can be fully characterized by the (simple) zeros $x_{n,j}$ of the
eigenfunction $u_n$ of $-\delx+V(x)$: $u_n(x_{n,j})=0$ for
$j=1,2,\ldots,n$ implies that $u(x_{n,j})=0$ for all eigenfunctions $u$ of
$H_n$. One thus can interpret $H_n$ as the strong coupling limit $g\to\infty$
limit of the following Hamiltonian, 
$$ 
H^{(g)}_{n} = -\delx^2 + V(x) + \sum_{j=1}^n g \, \delta( x-x_{n,j}
).
$$
Indeed, the eigenfunctions $u$ of this latter Hamiltonian are defined by
$-u''+Vu=Eu$ for $x\neq x_{n,j}$, together with the boundary conditions
$$ u(x_{n,j}+0) = u(x_{n,j}-0),\quad u'(x_{n,j}+0) - u'(x_{n,j}-0)=
gu(x_{n,j}+0) , 
$$ and for $g\to\infty$ the latter reduce to $u(x_{n,j})=0$. Thus the
factorization corresponds to adding infinitely strong repelling
delta-potentials at the locations of the zeros of the eigenfunction
$u_n$. This implies that the ground state of $H_n$ is $n+1$-fold
degenerate: all states $u_{0,j}(x)$ which are equal to $u_n(x)$ in an
interval between two adjacent zeros $x_{n,j}$ and zero otherwise are
eigenfunctions of $H_n$ with eigenvalue $E_n$. However, the excited
states cannot be found in such a simple manner in general.

\sct{5. Exactly solvable many-body systems in 1D}

\noindent We now show that the examples above allow a straightforward
extension to the $N$-particle case, and this gives rise to exactly
solvable interacting many body systems of {\em distinguishable}
particles.

\medskip
\noindent
{\bf 5.a.}  Interacting particles in a harmonic oscillator potential.

\smallskip
\noindent Our first example is for the operator
\begin{equation}
H_N =\sum_{j=1}^N (-\partial_{x_j}^2 + x_j^2 )  
\end{equation}
on $\R^N$. This self-adjoint operator is defined by the standard
factorization
$$ H_{N} = \sum_{j=1}^N (-\partial_{x_j}^2 + x_j^2 ) = \sum_{j=1}^N
\left(\partial_{x_j} - x_j \right)\left( -\partial_{x_j} - x_j \right)
+N,
$$ and it describes an arbitrary number, $N$, of noninteracting,
identical particles moving in an external harmonic oscillator
potential. The eigenfunctions of $H_{N}$ obviously are given by
products of the one particle eigenfunctions defined in Section~3.a,
$$
\psi_{\vn}(\vx) = \prod_{j=1}^N \psi_{n_j}(x_j),\quad
n_j=0,1,2,\ldots \quad , 
$$
where we use the shorthand notation $\vx=(x_1,\ldots,x_N)$ and similarly for
$\vn$, and the eigenvalue is
\begin{equation}
\cE_{\vn}=\sum_{j=1}^N (2n_j+1) .  \label{En} 
\end{equation}
Using the huge degeneracy of these eigenfunctions we can construct
eigenfunctions of $H_{N}$ which vanish at all coinciding points
$x_j=x_k$ by antisymmetrization,
\begin{equation} 
f_{\vn}(\vx) = \sum_{P\in S_N} (-1)^{|P|} \prod_{j=1}^N
\psi_{n_j}(x_{Pj}),\quad \quad 0\leq n_1<n_2<\ldots < n_N<\infty
\label{fnn}.  
\end{equation}
These are exactly the fermion eigenfunctions of $H_{N}$, of course:
exchanging any two particles gives a minus sign.  The fermion wave function
with the lowest eigenvalue is
\begin{equation}
f_0(\vx) = const\, \det_{1\leq j,k\leq N} \left(
\psi_{j-1}(x_k)\right) \label{f0}
\end{equation} 
where $const$ is some irrelevant non-zero constant which we will later
fix to some convenient value. To compute $f_0$ we use the explicit
form of the harmonic oscillator eigenfunctions in Eq.\ \Ref{psin1},
$$ f_{\vn_0}(\vx) = \ee^{-\sum_j x_j^2/2}\det_{1\leq j,k\leq N} (
\cH_{j-1}(x_k) ) .
$$ Since the Hermite polynomials are of the form $\cH_n(x) = c_n x^n +
\ldots $ with dots some linear combinations of $\cH_{m<n}$ and $c_n$
non-zero constants, we can replace in this determinant
$\cH_{j-1}(x_k)$ by $c_{j-1} x_k^{j-1}$, and a well-known identity of
the Vandermonde determinant implies that
$$
f_0(\vx) = 
\ee^{-\sum_{j=1}^N x_j^2/2} \prod_{1\leq j<k\leq N}(x_j-x_k) 
$$ is the fermion ground state of the operator $H_N$ with the
eigenvalue $\sum_{j=0}^{N-1} (2j+1) = N^2$ (we fixed $const$ so as to
absorb factors of $c_j$ and a possible sign). Then obviously
$$
\frac1{f_0(\vx)} \frac{\partial}{\partial x_j} f_0(\vx) =  - x_j + 
\mathop{\sum_{k=1}^N}_{k\neq j}
\frac1{x_j-x_k} . 
$$ We thus conclude that the differential operator defining $H_{N}$
also can be factorized as follows,
\begin{equation} 
H_{N,1} = \sum_{j=1}^N (-\partial_{x_j}^2 + x_j^2 ) = \sum_{j=1}^N
Q_{j}^* Q_{j}^\nd + N^2 ,\quad Q_{j} = -\partial_{x_j} - x_j +
\mathop{\sum_{k=1}^N}_{k\neq j}
\frac1{x_j-x_k} ,  \label{123} 
\end{equation}
and this defines another self-adjoint operator.

\noindent We now discuss how to determine all eigenfunctions $f(\vx)$
of $H_{N,1}$; see \cite{G} for a similar discussion. As in Section~2
we conclude that the eigenfunction have to vanish at coinciding
points,
\begin{equation}
f(x_1,x_2,\ldots,x_N) = 0 \quad \mbox{ if $x_j=x_k$ for some $j\neq k$,
$j,k=1,2,\ldots,N$} , \label{BC1} 
\end{equation}
and thus all fermion eigenfunctions $f_{\vn}$ of $H_{N}$ in Eq.\ \Ref{fnn}
are also eigenfunctions of $H_{N,1}$ with the same eigenvalues. It is not
difficult to see that all these are $C^\infty$ eigenfunctions of $H_{N,1}$,
but they obviously do not span the full Hilbert space $L^2(\R^N)$ but only its
fermion subspace. To find the missing eigenfunctions we divide the space
$\R^N$ in fundamental domains which can be parameterized by permutations $P\in
S_N$,
\begin{equation} 
\Delta_{P} : = \left\{ \vx=(x_1,\ldots,x_N)\in \R^N| -\infty <
x_{P1} < x_{P2} < \ldots < x_{PN} <\infty \right\} .  
\end{equation}
There are $N!$ distinct such domains, and the closure of the union of all
these domains equals to full space, $\overline{\cup_{P\in S_N}\Delta_P} =
\R^N$. Moreover, the boundaries of these domains correspond to the hyperplanes
$x_j=x_k$. We thus need to find all functions which are $C^\infty$ and obey
the eigenvalue equation for $H_{N}$ in the interior of all these domains
$\Delta_P$ and vanish at all their boundaries. The functions in Eq.\ \Ref{fnn}
obey both these conditions, but since the derivatives of eigenfunctions need
not be continuous at coinciding point also the functions
\begin{eqnarray} 
f^P_{\vn}(\vx) = \left\{ \begin{array}{cc} f_{\vn}(\vx) & \mbox{ for
$\vx\in\Delta_{P} $}\\ 0 & \mbox{ otherwise } \end{array} \right. ,
\quad 0\leq n_1<n_2<\ldots <n_N<\infty,\quad P\in S_N
\label{fP} 
\end{eqnarray}
are legitimate eigenfunctions: they obey the correct eigenvalue
equation with the eigenvalue in Eq.\ \Ref{En} at non-coinciding
points, and they also obey all boundary conditions in Eq.\ \Ref{BC1}.
It is not difficult to verify that these eigenfunctions $f_{\vn}(\vx)$
span the full Hilbert space $L^2(\R^N)$.

Similarly as discussed in Section~4, we can interpret $H_{N,1}$ as the
strong coupling limit, $g\to\infty$, of the following Hamiltonian
$$
H^{(g)}_N = \sum_{j=1}^N (-\partial_{x_j}^2 + x_j^2 ) + g\sum_{1\leq
j<k\leq N} \delta(x_j-x_k)
$$
describing $N$ particles moving on the real line in the harmonic
oscillator potentials and interacting with two-body delta function
potentials. This corresponds to the fact that the quadratic form of
the operator $H_{N,1}$ is split into $N!$ independent problems in
$L^2(\Delta_P)$ with Dirichlet boundary conditions.

\medskip
\noindent
{\bf 5.b.} Interacting particles in a P\"oschl--Teller potential.

\smallskip
\noindent 
We now consider the operator
\begin{equation}
H_N = \sum_{j=1}^N \left( -\partial_{x_j}^2
-\ga(\ga+1)\frac1{\cosh^2(x_j)} \right) \label{HN1}
\end{equation}
describing $N$ noninteracting particles in a weakly confining
$\cosh^{-2}$-potential. For $\ga>N$ the fermion ground state of this
Hamiltonian is $f_0$ as in \Ref{f0} but now with the eigenfunctions
$\psi_n$ in \Ref{PTn}, and by an argument as in Section~5.a we
conclude that
$$
f_0(\vx) = \prod_{j=1}^N \cosh^{-\ga}(x_j) \prod_{1\leq j<k\leq N}
\left( \sinh(x_j) -\sinh(x_k) \right)  
$$
with corresponding eigenvalue 
$$
\cE_{0} = - \sum_{n=0}^{N-1} (\ga-n)^2 = -\frac{N}6(2N-1)(N-1) +  \ga N(N-1) -N\ga^2 . 
$$
We thus can define the following Hamiltonian different from $H_N$, 
\begin{equation}
H_{N,1} = \sum_{j=1}^N Q_{j}^*Q_{j} +\cE_{0} \label{HN1b}
\end{equation}
with
\begin{equation} 
Q_{j} = -\partial_{x_j}  + \frac1{f_0} \partial_{x_j} f_0 = -\partial_{x_j}
- \ga\tanh(x_j) + \mathop{\sum_{k=1}^N}_{k\neq j}
\frac{\cosh(x_j)}{\sinh(x_j)-\sin(x_k)} .  \label{Qjb}
\end{equation}
As in Section~5.a we can interpret $H_{N,1}$ as the strong coupling
limit of the Hamiltonian $H_N$ in \Ref{HN1} with two body delta
interactions enforcing that the eigenfunctions vanish at coinciding
points $x_j=x_k$, and all eigenvalues and eigenfunctions of this
interacting many-body Hamiltonian can be determined explicitly from
the fermion wave functions and corresponding eigenvalues of the
non-interacting Hamiltonian $H_{N}$. The example here is, however,
quite interesting from a physical point of view since, in addition to
many-body bound states, there are also scattering states and, in
general, one has mixed states where only some of the particles are
trapped and the others are free.

\medskip
\noindent
{\bf 5.c.} Interacting particles on the circle.

\smallskip
\noindent 
Let 
\begin{equation} 
H_{N} = -\Delta = -\sum_{j=1}^N \partial_{x_j}^2 \label{Delta}
\end{equation}
on $N$ dimensional torus ${\mathbb T}^N$. The right hand side of
\eqref{Delta} can be considered as a trivial factorization of the
periodic Laplace operator. We note that 
$$ \psi_n(x) = (\ee^{\ii x})^n \ee^{-\ii (N-1)x/2},\quad n=0,\pm 1,\pm
2,\ldots 
$$ all are eigenfunctions of the corresponding one-particle
Hamiltonian $-\partial_x^2$ with eigenvalue $(n-\half (N-1))^2$, and
thus $f_0$ as defined in \Ref{f0} is an fermion eigenstate of the
operator \eqref{Delta} on ${\mathbb T}^N$ with the eigenvalue
$\sum_{n=0}^{N-1}(n-\half (N-1))^2 = \frac1{12}N(N^2-1)$. By
straightforward computations we find that
$$ f_0(\vx) = \prod_{1\leq j<k\leq N} \sin\half(x_j-x_k),
$$
and therefore, by analogy with \eqref{123}, we can consider the
factorization
\begin{equation} 
H_{N,1} = -\sum_{j=1}^N \partial_{x_j}^2 = \sum_{j=1}^N Q_{j}^*
Q_{j}^\nd + \frac{1}{12}N(N^2-1) \label{HN1c}
\end{equation}
where 
\begin{equation} 
Q_{j} = -\partial_{x_j} -\frac1{f_0} \partial_{x_j} f_0 =
-\partial_{x_j} + \mathop{\sum_{k=1}^N}_{k\neq j}
\half\cot\half(x_j-x_k). \label{Qjc} 
\end{equation}
Note that the singularities enforce that eigenfunctions of $H_{N,1}$
vanish at points $x_j=x_k$ modulo $2\pi$, and thus we can interpret
$H_{N,1}$ as describing particles moving on a circle of length $2\pi$,
$-\pi\leq x_j\leq \pi$, interacting with a strong two-body delta
interaction.

As in the previous examples this factorization splits the operator
$H_{N,1}$ into an orthogonal sum of operators in fundamental domains
parameterized by permutations $P\in S_N$, and the eigenfunctions of
$H_{N,1}$ can all be determined explicitly as antisymmetrized products
of the eigenfunctions of the corresponding non-interacting Hamiltonian
$H_{N}$.

\sct{6. A method to find Calogero-Sutherland type systems}

\noindent As already mentioned, our considerations in the previous
section are closely related to arguments which are well-known in the
context of Calogero-Sutherland type systems \cite{C0}. In this
sections we explore this relation in more detail. Through that we are
naturally lead to a surprisingly simple approach to a large class of
Calogero-Sutherland type systems.

\noindent One well-known method to find quantum integrable systems is
to write down a many-body state of Jastrow form
$$ \Psi_0(\vx) = \prod_{j=1}^N \ee^{-w(x_j)} \prod_{1\leq j<k\leq N}
g(x_j,x_k)^\lambda, \quad \lambda\geq 0,
$$ for some $C^2$ functions $w$ and $g$, and then construct a
many-particle Hamiltonian
$$H_N = \sum_{j=1}^N \cQ_j^* \cQ_j^{\phantom *}- \cE_0,\quad
\cQ_j=-\partial_{x_j} + \frac1{\Psi_0}\partial_{x_j}\Psi_0
$$ with $\cE_0$ some real, conveniently chosen constant. Then $H_N$
obviously defines self-adjoint operator $\geq \cE_0$, and
$\cQ_j\Psi_0=0$ $\forall j$ implies that $\Psi_0$ is an exact
groundstate of $H_N$ with eigenvalue $\cE_0$. In general, such a
Hamiltonian $H_N$ describes a system of particles with two- and
three-body interactions, and only for very particular functions $w$
and $g$ the three-body terms cancel. These cases are of particular
interest since they lead to physically interesting Hamiltonians of the
form as in Eq.\ \Ref{HN} which, in many cases, are exactly
solvable. Famous such examples are $w(x)=\half x^2$, $g(x,y)=(x-y)$
and $w(x)=0$, $g(x,y)=\sin\half(x-y)$ corresponding to the models
discovered by Calogero \cite{C}:
$$ H_N = \sum_{j=1}^N \Bigl( -\partial_{x_j}^2 + x_j^2 \Bigr) +
\lambda(\lambda-1) \sum_{1\leq j<k\leq N} \frac2{(x-y)^2}
$$ and Sutherland \cite{S}: 
$$ H_N = -\sum_{j=1}^N \partial_{x_j}^2 + \lambda(\lambda-1)
\sum_{1\leq j<k\leq N} \frac1{2\sin^2\half(x_j-x_j)},
$$ respectively. Olshanetsky and Perelomov found interesting variants
of the Calogero-Sutherland models related to root systems; see
\cite{OP} and references therein.

Obviously, the Calogero- and Sutherland models are very closely
related to our examples discussed in Sections~5.a and 5.c,
respectively. It is interesting to note that our considerations there
provide an alternative proof of the key identity leading to these
models. Indeed, if we check the identity in (\ref{123}) by a direct
computation, we find that it is equivalent to the well-known functional
identity which cancels the three-body terms in the Calogero model:
$$
\frac1{(x_j-x_k)(x_j-x_\ell)} + cycl = 0 
$$
for $j,k,\ell$ all different, where `$+cycl$' is short for adding the
two terms of the same form obtained by cyclic permutations of the
indices, $(j,k,\ell)\to (k,\ell,j)$ and $(j,k,\ell)\to (\ell,j,k)$.
However, since we have a direct proof of (\ref{123}) based on a basic
identity of the Vandermonde determinant, we found an alternative proof
of this functional identity. Similarly, the identity in Eqs.\
\Ref{HN1c}--\Ref{Qjc} is equivalent to the well-known functional
identity
$$ \cot(\half(x_j-x_k)) \, \cot\half(x_j-x_\ell) + cycl = -1, 
$$ used to cancel the three-body terms in the Sutherland model.

It therefore is natural to ask: is there a Calogero-Sutherland type
system corresponding to our example in Section~5.b? We now can use a
similar method as above to {\em find} such a system: we can use Eqs.\
(\ref{HN1b})--(\ref{Qjb}) to derive the functional identity
$$ \frac{\cosh^2(x_j)}{\left(\sinh(x_j)-\sinh(x_k)\right)
\left(\sinh(x_j)-\sinh(x_\ell)\right)} + cycl = 1,
$$ which then can be used to cancel three-body terms and prove that
\begin{equation} 
\Psi_0(\vx) = \prod_{j=1}^N \cosh^{-\ga}(x_j) \prod_{1\leq j<k\leq
N} \left( \sinh(x_j) -\sinh(x_k) \right)^\lambda 
\end{equation}
is the exact ground state of the Hamiltonian
\begin{equation} 
H_N = \sum_{j=1}^N \left( -\partial_{x_j}^2
-\frac{\ga(\ga+1)}{\cosh^{2}(x_j)} \right) +\lambda(\lambda-1)
\sum_{1\leq j<k\leq N}
\frac{\cosh^2(x_j)+\cosh^2(x_k)}{\left(\sinh(x_j)-\sinh(x_k)\right)^2}
.
\label{HCSb} 
\end{equation}
Moreover, it is straightforward to show that all other bound states
and corresponding eigenvalues can be found using Sutherland's method
\cite{S}.\footnote{E. Langmann, unpublished.}

It is interesting to note that our argument can be easily extended to
a rather large class of examples: assume a one-particle Hamiltonian
$H=-\partial_x^2+V(x)$ which has exact eigenstates of the form 
$$\psi_n(x)= P_n(\eta(x))\ee^{-w(x)},\quad n=0,1,2,\ldots
$$ with eigenvalue $E_n$, where $P_n$ is a polynomial of order $n$ and
$w$ and $\eta$ are $C^2$ functions.  Then $f_0$ in \Ref{f0} is
obviously an eigenstate of the many-body Hamiltonian
$$\sum_j(-\partial_{x_j}^2 + V(x_j))$$ with eigenvalue $\cE_{0} =
\sum_{j=0}^{N-1} E_n$.  As in Section~5 we can conclude that
$$f_0(\vx) = \prod_{j=1}^N \ee^{-w(x_j)} \prod_{1\leq j<k\leq N}
\left( \eta(x_j)-\eta(x_k)\right), 
$$ and using this formula for $f_0$ the identity
$$\sum_{j=1}^N (-\partial_{x_j}^2 + V(x_j)) = \sum_{j=1} Q_{j}^*
Q_{j} + \cE_{0},\quad Q_{j}=-\partial_{x_j} +
\frac1{f_0}\partial_{x_j} f_0$$ implies a functional identity
expressing three-body terms through two-body terms.
We now compute
$$ \cH = \sum_{j=1} \cQ_{j}^*\cQ_{j},\quad \cQ_{j}=-\partial_{x_j} +
\frac1{\Psi_0}\partial_{x_j} \Psi_0
$$ with $\Psi_0$ in \Ref{Psi0}, and using this functional identity we
can cancel all three-body terms and thus find that $\cH=H_N-\cE_0$
with $H_N$ in \Ref{HN}--\Ref{W} and $\cE_0$ in \Ref{cE0}.  Since
obviously $\cH\Psi_0=0$ and $\cH\geq 0$, we obtain the result stated
in the Proposition in Section~1. Table~\ref{table1} gives other
interesting examples of this result. We plan to give a comprehensive
discussion of all these models and their solution in a future
publication \cite{HL3}.

\sct{7. Final comments}

\noindent
We should mention that the model in \Ref{HCSb} is closely related to
the well-known $C_N$-variant of the Sutherland model defined by
\begin{eqnarray*}
H_{C_N} = \sum_{j=1}^N \left(-\partial_{x_j}^2
+\frac{\gamma(\gamma+1)}{\sinh^2(x_j)} \right) +  \lambda(\lambda-1)
\sum_{1\leq j<k\leq N} \biggl( \frac1{2\sin^2\half(x_j-x_j)} +\\ 
\frac1{2\sin^2\half(x_j+x_j)} \biggr);
\end{eqnarray*}
see \cite{OP} and references therein. This Hamiltonian describes
identical particles in an external, repulsive potential and with
repulsive two-body interactions, and it only has scattering states
(purely continuous spectrum). One can show that by shifting $x_j\to
x_j + \ii \pi/2$, this Hamiltonian is converted to the one in
\Ref{HCSb}, up to an additive constant. However, these shifts changes
the physics of the model dramatically: the Hamiltonian in \Ref{HCSb}
describes interacting particles in a weakly confining potential, and
it thus has, in addition to bound states and scattering states, also
mixed states where some of the particles are confined and others free.
To our knowledge, this situation has not been studied in the context
of Calogero-Sutherland type systems.  We therefore think it would be
interesting to study this model in detail which, however, is beyond
the scope of the present paper.

\noindent
There are other special cases which seem less known in the physics
literature even though, to our opinion, they also of interest in
physics. The examples listed in Table~\ref{table1} lead to systems
which were previously known in the mathematics literature; see
\cite{BF,DX} and references therein. We should mention that these
latter systems are closely related to the $BC_N$ variants of the
Calogero- (IV) and Sutherland model (I--III) \cite{OP}, respectively.
However, our result nevertheless seems useful since it is the starting
point of an alternative, systematic and unified approach to these
systems and their solutions, including examples which are hardly known
\cite{HL3}.

\noindent {\bf Acknowledgement} We would like to thank G\"oran
Lindblad for providing with us a helpful tool for numerical
experiments \cite{GL}. We are grateful for helpful discussions with
F.\ Calogero, P.\ Forrester, M.\ Halln\"as, G.\ Lindblad and J.\
Mickelsson. C.P. was supported by {\em Deutsche
Forschungsgemeinschaft} under the Emmy-Noether program. E.L. was
supported by the Swedish Science Research Council~(VR). E.L. and
A.L. are supported by the G\"oran Gustafsson Foundation and the
European grant ``ENIGMA''.

\end{document}